\documentclass[aps,preprint,superscriptaddress,amsmath,amssymb,nofootinbib]{revtex4-1}
\usepackage{amsmath}
\usepackage{amssymb}
\usepackage{amsfonts}
\usepackage{graphicx}
\usepackage{color}
\usepackage[stable]{footmisc}
\usepackage{IEEEtrantools}
\usepackage{float}
\usepackage{caption}
\usepackage{subcaption}
\numberwithin{equation}{section}

\numberwithin{equation}{section}

\restylefloat{figure}

\begin{document}

\title{Thin $\theta$-films optics}

\author{Luis Huerta}
\email{lhuerta@utalca.cl}
\homepage{http://www.pppp.cl}
\affiliation{Facultad de Ingenier\'{\i}a, Universidad de Talca, 3340000 Curic\'o, Chile.}
\affiliation{{\rm P}$^4$-Center for Research and Applications in Plasma Physics and Pulsed Power Technology, 7600713 Santiago, Chile.}

\begin{abstract}

\noindent
A Chern-Simons theory in 3D is accomplished by the so-called $\theta$-term in the action, $(\theta/2)\int F\wedge F$, which contributes only to observable effects on the boundaries of such a system. When electromagnetic radiation interacts with the system, the wave is reflected and its polarization is rotated at the interface, even when both the $\theta$-system and the environment are pure vacuum.  These topics have been studied extensively. Here, we investigate the optical properties of a thin $\theta$-film, where multiple internal reflections could interfere coherently.  The cases of pure vacuum and a material with magneto-electric properties are analyzed. It is found that the film reflectance is enhanced compared to ordinary non-$\theta$ systems and the interplay between magneto-electric properties and $\theta$ parameter yield film opacity and polarization properties which could be interesting in the case of topological insulators, among other topological systems. \\

\end{abstract}

\maketitle

\section{INTRODUCTION}

Chern-Simons (CS) theories have found applications in various areas, from gauge field theories and three-dimensional gravity \cite{DJT,3DGravity} to condensed matter physics.  In the latter case, CS formalism has been applied to phenomena that exhibit topological properties, a remarkable example of that being the quantum Hall effect, occuring in two spatial dimensions (2D) \cite{QHE}  (see also \cite{Z2005}). But, also in 3D space, CS theories have become relevant. The so-called $\theta$-term introduced in QCD by Peccei and Quinn, and generalized by Wilczek, represent the 3+1 spacetime dimension version of CS forms.\cite{Peccei1977,Wilczek87}. In recent years, the novel 3D topological insulators have become a new example of the above, and intensive research has been developed in the field \cite{TI2008,TI2011}. 

A 3D CS theory is characterized by the term $(\theta/2)\int F\wedge F$ in the action, which is a border term and it is relevant only on the system boundaries. As a consequence, electromagnetic radiation inciding on the interface between an ordinary medium and a $\theta$-system, matter or even pure vacuum, exhibits a polarization rotation both for reflected and refracted waves.  The effect is modulated by the interplay between magnetic and dielectric properties of the system and the value of the $\theta$ parameter. This phenomenon has been reported as the Kerr-Faraday rotation present in topological insulators \cite{Tse-McD2010} (see Ref.~\cite{TI2008} for a review). Apart from the effect of polarization, the interface reflectance increases compared to ordinary systems.  In a previous paper we reported in detail the optical properties of a $\theta$-material interface \cite{Huerta2014}. The above phenomena are even stronger in the case of pure vacuum, where the only difference between the $\theta$-vacuum and the surrounding medium is a nonzero value for $\theta$ inside the system \cite{HZ2012}. At both sides of the interface, however, electromagnetic propagation is the same.
 
Although the $\theta$-vacuum case is the clearest manifestation of the peculiarity involved in these systems, applications to material systems are interesting and useful, since nontrivial topology is an extended issue in physics. In particular, the study of thin $\theta$-films properties could be relevant for applications, since they are often seen in such a geometry \cite{Jiang2012}. We study here the optical properties of a generic thin $\theta$-film, surrounded by ordinary matter or vacuum. We also include the possibility of pure vacuum in both the film and the medium outside.  In particular, we investigate theoretically the system behavior under electromagnetic radiation. Since in (3+1)D, CS term becomes a boundary term, it will not contribute to field equations in the bulk of a system, but will only affect the system at its boundaries. 

For thin films, interference effects are interesting, those resulting from the multiple coherent superposition of waves emerging on both sides of the film, and that is a relevant topic in the present paper. This report is organized as follows. First, we present the basics of the theory and recall some results for a single interface, from Ref.~\cite{Huerta2014}. In the following sections we develop our approach to the film problem and present our findings.

\section{THE $\theta$-SYSTEM INTERFACE}

For a $\theta$-system interacting with electromagnetic fields, as in the case of radiation, the action is written as $S_{M} + \theta \int F\wedge F$ ($S_M$ is the Maxwell action), and field equations are modified only at the boundaries. We denote the boundary surface by $\Sigma$, with a locally defined normal unit vector $\bf{\hat n}$ pointing inside the system. In terms of noncovariant electric and magnetic fields, the field equations are \cite{HZ2010}
\begin{eqnarray}
\varepsilon \nabla  \cdot {\bf{E}} &=& \theta {\kern 1pt} \delta \left( \Sigma  \right){\bf{B}} \cdot {\bf{\hat n}} \label{Emat_eq} \\
\frac{1}{\mu} \nabla  \times {\bf{B}} - \partial_t {\bf E} &=& \theta {\kern 1pt} \delta \left( \Sigma  \right){\bf{E}} \times {\bf{\hat n}} , \label{Bmat_eq}
\end{eqnarray}
where $\varepsilon$ and $\mu$ represent the electric permittivity and magnetic permeability of the material, respectively.\footnote{We choose $c=1$. Consequently, the vacuum magnetoelectric properties are $\varepsilon_0=\mu_0=1$, and the system's $\varepsilon$, $\mu$ are dimensionless.}  The delta function in the RHS of the above equations stands for restricting the effects to the surface, with the corresponding terms representing the surface charge and current densities built from the electromagnetic field itself. The set of equations is completed with the homogeneous Maxwell equations $\mathbf{\nabla }\cdot \mathbf{B}=0$ and $\nabla \times E + \partial_t {\bf B}=0$. By the standard procedure, we obtain a pair of discontinuity conditions at the interface $\Sigma$:
\begin{eqnarray}
\left[ \varepsilon {\bf E}_{\rm n} \right] &=& \theta {\bf{B}}_{\rm n} \label{Emat_discon} \\
\left[ \dfrac{1}{\mu} {\bf B}_\tau \right]  &=&  - \theta {\bf E}_\tau . \label{Bmat_discon}
\end{eqnarray} 
Subindices ${\rm n}$ and $\tau$ stand for normal and tangent components of the fields, respectively, with respect to the interface $\Sigma$. The symbol $\left[ \right]$ must be interpreted as the difference between the fields evaluated immediately inside and immediately outside the $\theta$-material. Additionally, we have, from the homogeneous equations, the standard continuity of ${\bf{E}}_{\tau}$ and ${\bf{B}}_{\rm n}$ components.

We consider a linearly polarized planar electromagnetic wave inciding on the boundary surface of a $\theta$-system, with an angle $\varphi$ with respect to the normal. The standard laws of reflection and refraction hold, independently of the polarization, and the transmitted wave into the $\theta$-medium is refracted with an angle $\psi$.  By applying the boundary conditions derived above, we find the electromagnetic field amplitudes for the transmitted (refracted) and reflected waves. We adopt the usual decomposition of the fields in parallel ($\parallel$, ${\rm p}$) and perpendicular  ($\bot$, ${\rm s}$) components with respect to the incidence plane, and introduce the adimensional field amplitudes in terms of the incident wave field $E_i$: ${e_{i\parallel }} \equiv {E_{i\parallel }}/{E_i}$, ${e_{r\parallel }} \equiv {E_{r\parallel }}/{E_i}$ and ${e_{t\parallel }} \equiv {E_{t\parallel }}/{E_i}$, with ${E_i} = {({{E_{i\parallel }}^2 + {E_{i \bot }}^2})^{1/2}}$, and similar definitions  for the corresponding ${\rm s}$ components.\footnote{As is known, at each medium magnetic and electric fields are related one to each other by ${\mathbf{B}} = n{\bf{\hat k}} \times {\mathbf{E}} \label{E_B}$, where $n = {( \mu \varepsilon )^{1/2}}$ is the respective refraction index of the medium and $\hat{\bf k}$ is the corresponding unit wave vector.} Then, for the transmitted wave, we find \cite{Huerta2014}
\begin{equation}
\left( {\begin{array}{*{20}{c}}
{{e_{t\parallel }}}\\
{{e_{t \bot }}}
\end{array}} \right) = \frac{2}{D}\left( {\begin{array}{*{20}{c}}
{\eta s + 1}&{ - \theta }\\
{\theta s}&{\eta  + s}
\end{array}} \right)\left( {\begin{array}{*{20}{c}}
{{e_{i\parallel }}}\\
{{e_{i \bot }}}
\end{array}} \right) \label{e_t}
\end{equation}
and, for the reflected wave,
\begin{equation}
\left( {\begin{array}{*{20}{c}}
{{e_{r\parallel }}}\\
{{e_{r \bot }}}
\end{array}} \right) = \frac{1}{D}\left( {\begin{array}{*{20}{c}}
{\left( {\eta s + 1} \right)\left( {\eta  - s} \right) + {\theta ^2}s}&{2\theta s}\\
{2\theta s}&{ - \left( {\eta s - 1} \right)\left( {\eta  + s} \right) - {\theta ^2}s}
\end{array}} \right)\left( {\begin{array}{*{20}{c}}
{{e_{i\parallel }}}\\
{{e_{i \bot }}}
\end{array}} \right) \label{e_r},
\end{equation}
where $D \equiv \left( {\eta s + 1} \right)\left( {\eta  + s} \right) + { \theta^2}s$. We have introduced the convenient definitions $s \equiv \cos \psi /\cos \varphi$, and $\eta  \equiv (n_2/n_1)(\mu_1/\mu_2) = \sqrt{ (\varepsilon_2\mu_1)/(\varepsilon_1\mu_2) } $, which describes the dielectric and magnetic properties of the system. Also, to simplify notation we have redefined $\theta / {\eta _1} \to \theta$.\footnote{If the surrounding medium is vacuum space, the redefined $\theta$ coincides with the parameter in the action.}

Unlike the normal systems, for $\theta$-systems the ${\rm p}$ and ${\rm s}$ components of the reflected and refracted waves mix each other when crossing the system interface.  Thus, both waves experience changes in their polarization. We find, for the transmitted wave, a polarization angle, measured with respect to the incidence plane, given by\footnote{The right-hand rule is followed in the definition of the polarization angle, with the corresponding wave vector $\hat{\bf k}$ as the reference axis.}
\begin{equation}
{\alpha _{\scriptscriptstyle T}} = \tan^{-1} \left[ \frac{{\theta s + \left( {\eta  + s} \right)\tan {\alpha _{\scriptscriptstyle I}}}}{{\eta s + 1 - \theta \tan {\alpha _{\scriptscriptstyle I}}}}\right] \label{tPol}
\end{equation}
and, for the reflected wave,
\begin{equation}
{\alpha _{\scriptscriptstyle R}} = \tan^{-1} \left[ \frac{{2\theta s - \left[ {\left( {\eta s - 1} \right)\left( {\eta  + s} \right) + {\theta ^2}s} \right]\tan {\alpha _{\scriptscriptstyle I}}}}{{\left( {\eta s + 1} \right)\left( {\eta  - s} \right) + {\theta ^2}s + 2\theta s\tan {\alpha _{\scriptscriptstyle I}}}} \right] , \label{rPol}
\end{equation}
with $\alpha_{\scriptscriptstyle I}$ representing the incident wave polarization angle.

\vspace{0.5cm}
\begin{figure}[h]
        \centering
        \begin{subfigure}[b]{0.5\textwidth}
                \includegraphics[width=\textwidth]{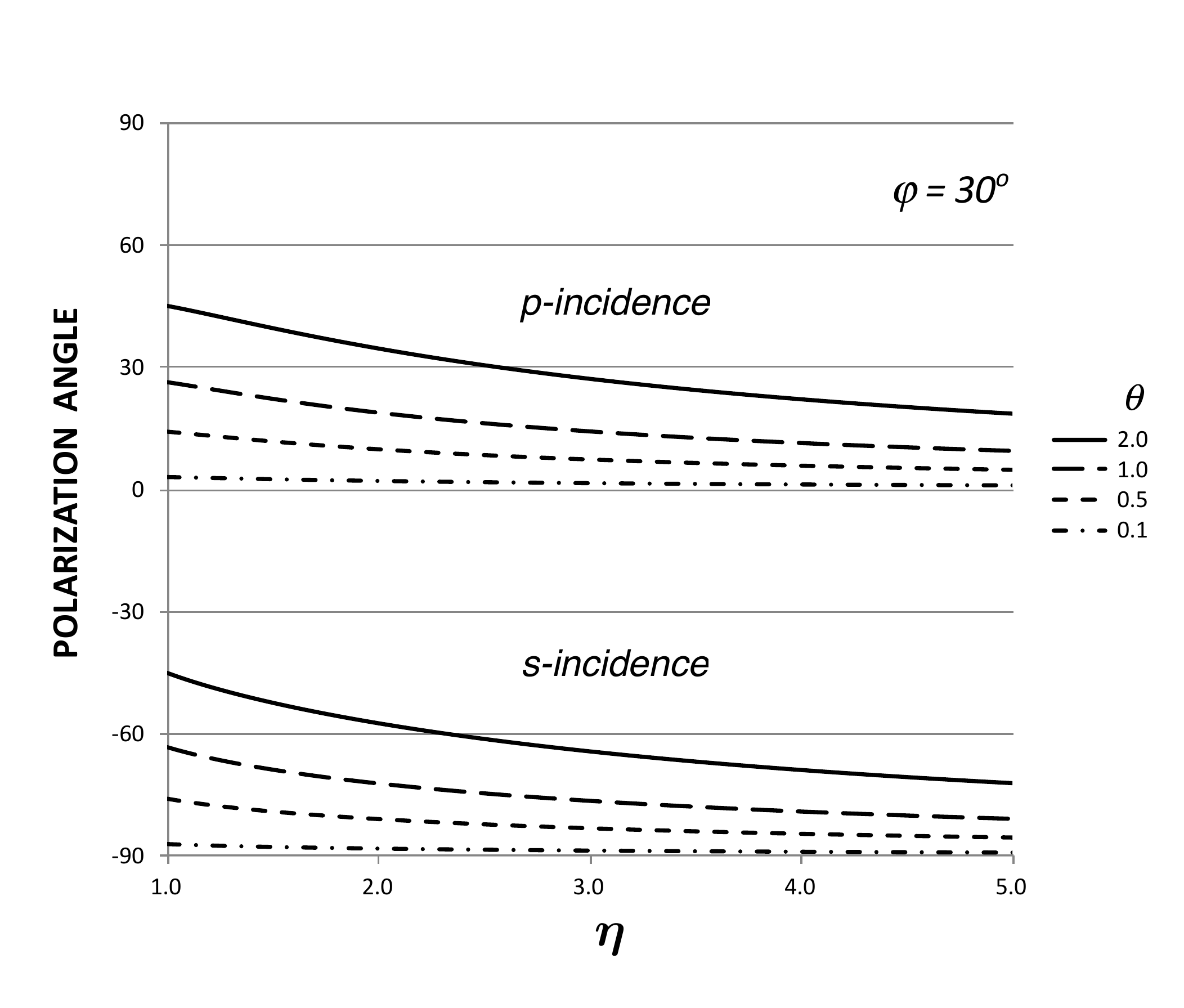}
                \caption{}
                \label{fig_tPol-eta}
        \end{subfigure}%
        ~ 
        \begin{subfigure}[b]{0.5\textwidth}
                \includegraphics[width=\textwidth]{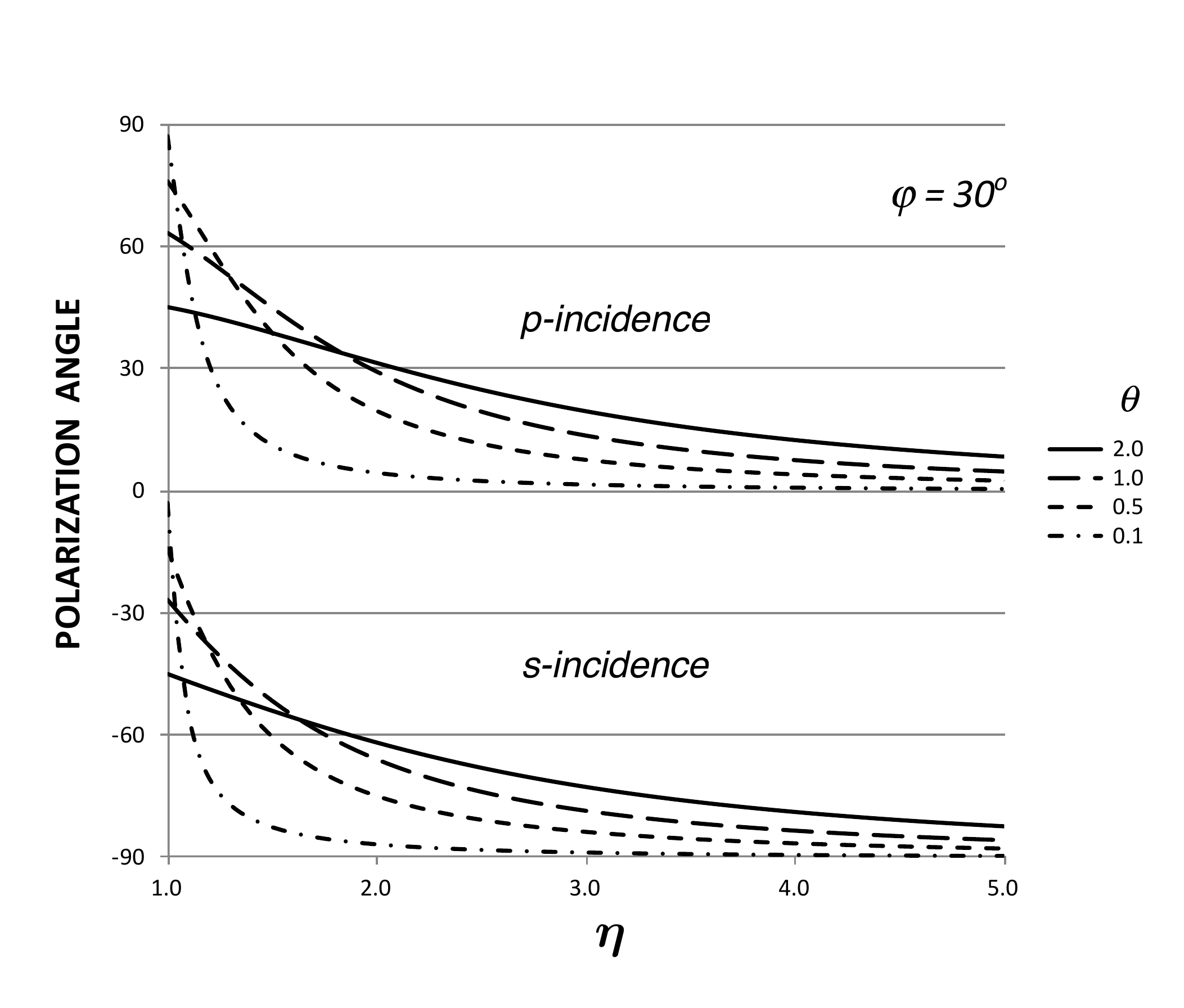}
                \caption{}
                \label{fig_rPol-eta}
        \end{subfigure}
        ~ 
               \caption{Polarization of transmitted (left) and reflected (right) waves by a single $\theta$-interface. Although for ordinary materials ($\theta=0$) there is no polarization rotation, for rather small nonvanishing values of $\theta$, a significant change in polarization is found which is larger for smaller magnetoelectric properties. Curves for p and s incidence look similar, but there are slight differences in the actual values of the respective polarization angles.}
               \label{fig_Pol-eta}
\end{figure}

Figure \ref{fig_Pol-eta} shows the polarization angle versus $\theta$ for different values of the magnetoelectric parameter $\eta$ (for ${\rm p}$-polarized incident wave). Included also is the $\theta$-vacuum ($\eta=1$) case for comparison. Because of ${\rm p}$-polarization incidence ($\alpha_{\scriptscriptstyle I} =0$), the curves represent the effective polarization rotation experienced by the outgoing waves. The interplay between magnetoelectric properties, $\eta$, and the $\theta$ parameter unveils different qualitative situations. For the transmitted wave to the material, $\eta$ diminishes the effect (except for values close to 1) with respect to the $\theta$-vacuum. For reflected radiation the situation is more interesting, but, similar to transmission, only for values of $\eta$ close to 1, there is an enhancement in polarization rotation  For large $\eta$ (not shown in Fig.~\ref{fig_Pol-eta}), only for very large values of $\theta$, for the transmitted wave, or $\theta$ large enough, for the reflected wave, there will be a significant polarization rotation. An example of this is the existence of a maximum for reflected wave polarization for $\eta > s$. A more detailed analysis is found in Ref.~\cite{Huerta2014}.

\section{$\theta$-FILM OPTICS}

\begin{figure}[h]
\begin{center}
\includegraphics[scale=0.5]{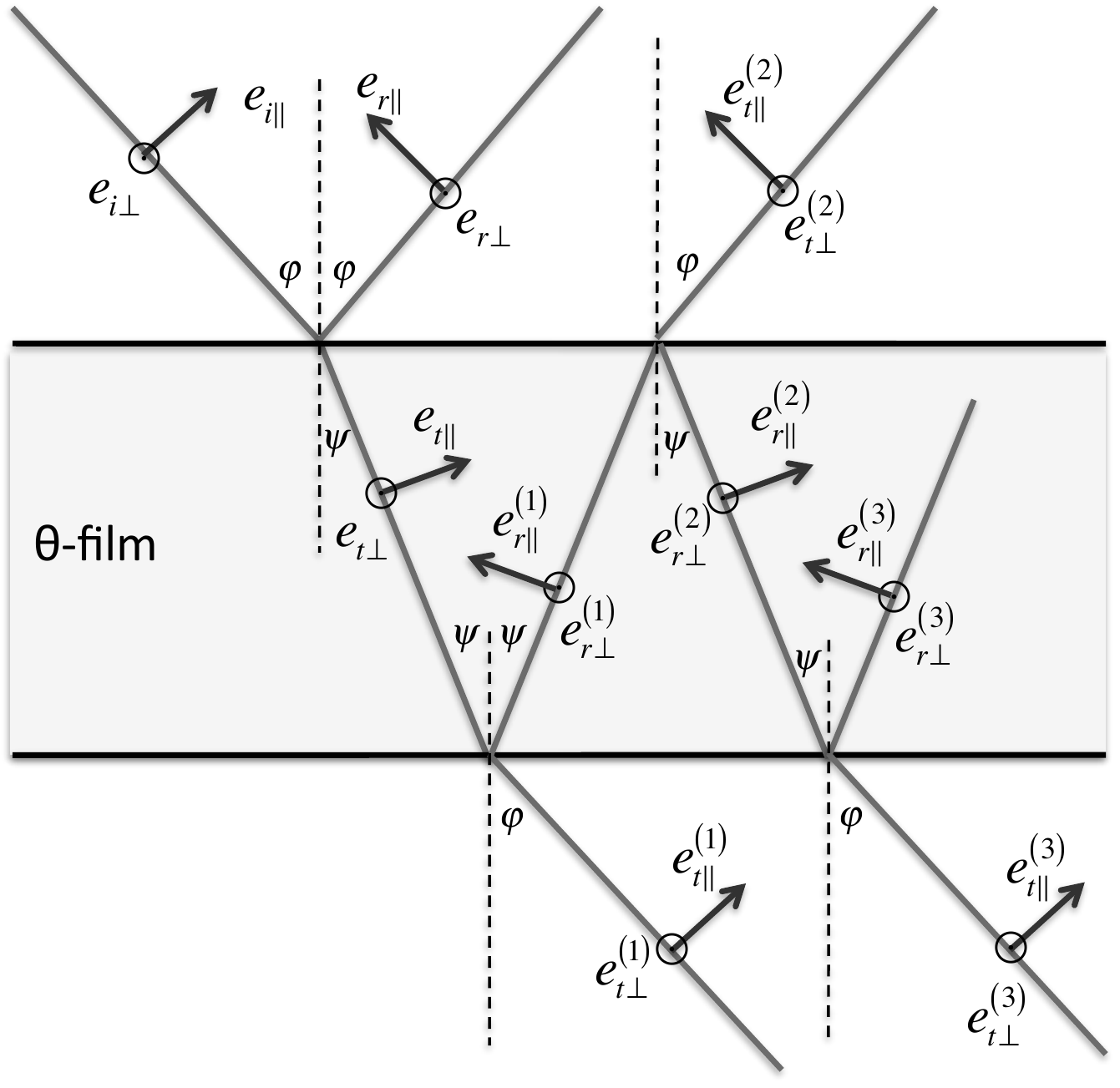}
\caption{Film geometrical optics. }
\label{fig-film}
\end{center}
\end{figure}

Let us now consider an infinite rectangular thin film of a $\theta$-system immersed in a normal non-$\theta$ medium. Both the film and the surrounding medium have different dielectric and magnetic properties, so that $\eta\ne 1$. When the electromagnetic wave reaches the film, it experiences a series of internal reflections between the two interfaces that separate the film from the medium outside. After each of those internal reflections, a wave emerges outside the film, from each side alternatively.  Fig.~\ref{fig-film} illustrates the geometry and definitions we use here.

Internally reflected waves yield refracted and reflected waves with field amplitudes expressed in terms of the arriving wave field. The discontinuity at the interface for internal waves is given by (\ref{Emat_discon}) and ((\ref{Bmat_discon}), but provided the substitution $\theta  \to  - \theta$. We define 
\begin{equation}
{\cal T} \equiv \frac{{2\eta s}}{D}\left( {\begin{array}{*{20}{c}}
{\eta s + 1}&{\theta s}\\
{ - \theta }&{\eta  + s}
\end{array}} \right) \label{T_matrix}
\end{equation}
as the matrix which represents the transformation of field amplitudes when the wave goes out from the film (in any boundary). For internal reflections, we have the transformation matrix
\begin{equation}
{\cal R} \equiv \frac{1}{D}\left( {\begin{array}{*{20}{c}}
{{\theta ^2}s - \left( {\eta s + 1} \right)\left( {\eta  - s} \right)}&{ - 2\theta \eta s}\\
{ - 2\theta \eta s}&{ - {\theta ^2}s + \left( {\eta s - 1} \right)\left( {\eta  + s} \right)}
\end{array}} \right) . \label{T_matrix}
\end{equation}
It is readily seen that the transmitted and reflected wave amplitudes are given by
\begin{equation}
e_t^{(l)} = {\cal T}{{\cal R}^{l - 1}}{e_t} , \label{et_l-0}
\end{equation}
\begin{equation}
e_r^{(l)} = {{\cal R}^l}{e_t} , \label{er_l-0}
\end{equation}
for $l = 1,\,2,\,...$ ($e_t$ is the field amplitude of the wave refracted at the incident interface).  The number $l$ represent the $l$th time that a film interface is reached by the internal traveling wave, with field amplitudes $e_t^{(l)}$ and $e_r^{(l)}$ for the transmitted and reflected waves, respectively. We note that $e_t^{(l)}$, for $l$ even, represent the field amplitudes of the waves reflected back by the film. Of course, for $l$ odd, we have the radiation waves passing through the film.

\section{COHERENT SUPERPOSITION OF WAVES}

If the film is thin enough, compared to the radiation wavelength, coherent waves will interfere with each other, for both the transmitted and reflected beams going out from the film. Each wave carries a phase shift $e^{i2n\bar k{\kern 1pt} \bar d}$, for the optical path back and forth inside the film. There, $n = {n_2}/{n_1}$ is the relative refraction index of the film with respect to the surrounding medium, $2n\bar k{\kern 1pt} \bar d$ thus representing the round optical path across the film. The corresponding wave number is ${n_2}k = n\bar k$, with $\bar k \equiv {n_1}k$, ${n_1}$ being the refraction index for the medium outside and $k$, the wave number in empty space. The distance $\bar d \equiv d/\cos \psi  = d/s\cos \varphi$ is the effective path length inside the film, which depends on the angle of incidence $\varphi$ ($d$ is the film thickness). No additional phase shift is introduced in the internal reflections, since we assume that $n_2 > n_1$.

For the radiation passing through the film, the resulting field amplitude ${{\cal E}_t}$ becomes  
\begin{equation}
{{\cal E}_t} = {\cal T}\left( {1 + {{\cal R}^2}{e^{i2n\bar k{\kern 1pt} \bar d}} + {{\cal R}^4}{e^{i4n\bar k{\kern 1pt} \bar d}} + ...} \right){e_t} , 
\label{Et-matrix}
\end{equation}
and, for the radiation reflected back, the field amplitude is
\begin{equation}{{\cal E}_r} = {e_r} + {\cal T}{\cal R}{e^{i2n\bar k{\kern 1pt} \bar d}}\left( {1 + {{\cal R}^2}{e^{i2n\bar k{\kern 1pt} \bar d}} + ...} \right){e_t}{\kern 1pt} .  \label{Er-matrix}
\end{equation}
The geometric series, represented by the matrix
\begin{equation}
{\cal M} \equiv 1 + {{\cal R}^2}{e^{i2n\bar k{\kern 1pt} \bar d}} + {{\cal R}^4}{e^{i4n\bar k{\kern 1pt} \bar d}} + ...  ,
\end{equation}
can be summed up by diagonalizing the reflection matrix $\cal R$, to obtain
\begin{equation}
{\cal M} = U\left( {\begin{array}{*{20}{c}}
{\dfrac{1}{{1 - {\lambda _ + }^2{e^{i2n\bar k{\kern 1pt} \bar d}}}}}&0\\
0&{\dfrac{1}{{1 - {\lambda _ - }^2{e^{i2n\bar k{\kern 1pt} \bar d}}}}}
\end{array}} \right){U^{ - 1}} .
\end{equation}
$\lambda_{\pm}$ are the eigenvalues of ${\cal R}$, given by
\begin{equation}
{\lambda _ \pm } = \frac{\eta }{D}\left( {{s^2} - 1} \right) \pm \frac{1}{D}Q ,
\end{equation}
and $U$ is the corresponding unitary transformation matrix in the diagonalization process:
\begin{equation}
U = \frac{1}{{\sqrt {2Q} }}\left( {\begin{array}{*{20}{c}}
{\sqrt {\delta  + Q} }&{\sqrt { - \delta  + Q} }\\
{ - \sqrt { - \delta  + Q} }&{\sqrt {\delta  + Q} }
\end{array}} \right) ,
\end{equation}
where $Q \equiv \sqrt{( {D - \eta {{\left( {s - 1} \right)}^2}} ) ( {D - \eta {{\left( {s + 1} \right)}^2}} )}$ and $\delta  \equiv {\theta ^2}s - s({{\eta ^2} - 1})$.

\section{REFLECTANCE FOR COHERENT RADIATION}

For constructive interference, $2n\bar k{\kern 1pt} \bar d = 2\pi$. Therefore, the field amplitude of radiation transmitted through the film is given by
\begin{equation}
{\cal E}_t^{\left( {\rm C}\right)} = {\cal T}{{\cal M}^{\left( {\rm C}\right)}}{e_t}  
 = {e_i} \, .
\end{equation}
Consequently,
\begin{equation}
{{\cal E}_r^{\left( {\rm C}\right)}} = 0 ,
\end{equation}
and the film becomes completely transparent, without any effect in the radiation polarization.

For destructive interference, $2n\bar k{\kern 1pt} \bar d = \pi$, and we obtain for the transmitted radiation through the film
\begin{equation}
{{\cal E}_t}^{\left( {\rm D} \right)} =
\frac{{2\eta }}{\Lambda }\left( {\begin{array}{*{20}{c}}
{{s^2}\left( {D - 2\eta } \right) - s\left( {{s^2} - 1} \right)\left( {\eta s + 1} \right)}&{\theta s\left( {{s^2} - 1} \right)}\\
{\theta s\left( {{s^2} - 1} \right)}&{\left( {D - 2\eta {s^2}} \right) + \left( {{s^2} - 1} \right)\left( {\eta  + s} \right)}
\end{array}} \right) \left( {{e_i}} \right)  .  \label{Et_des}
\end{equation}

For the radiation reflected by the film, the field amplitude is given by
\begin{equation}
{{\cal E}_r}^{\left( {\rm D} \right)} = \dfrac{1}{\Lambda }\left( {\begin{smallmatrix}
{{D^2} - 2\left( {s\left( {\eta s + 1} \right) + \eta } \right)D + 2\eta \left( {\eta  + s} \right)\left( {{s^2} + 1} \right)}&{2\theta s\left[ {D - \eta \left( {{s^2} + 1} \right)} \right]}\\
{2\theta s\left[ {D - \eta \left( {{s^2} + 1} \right)} \right]}&{ - {D^2} + 2\left[ {s\left( {\eta s + 1} \right) + \eta } \right]D - 2\eta s\left( {\eta s + 1} \right)\left( {{s^2} + 1} \right)}
\end{smallmatrix}} \right)\left( {{e_i}} \right) , \\*
\end{equation}
with $\Lambda  \equiv {[ {D - \eta ( {{s^2} + 1} )} ]^2} + {\eta ^2}{( {{s^2} - 1} )^2}$.

Therefore, the film transmittance $T \equiv {\left| {{{\cal E}_t}^{\left( {\rm D} \right)}} \right|^2}$ and reflectance $R \equiv {\left| {{{\cal E}_r}^{\left( {\rm D} \right)}} \right|^2}$, computed for p incidence, are, respectively,
\begin{equation}
T^{\left( {\rm D} \right)}  = \frac{{4{\eta ^2}{s^2}}}{{{\Lambda ^2}}}\left[ {{{\left( {{\theta ^2}{s^2} + {\eta ^2}{s^2} + 1} \right)}^2} + {\theta ^2}{{\left( {{s^2} - 1} \right)}^2}} \right] ,  \label{T_(D)}
\end{equation}
\begin{equation}
R^{\left( {\rm D} \right)}  = \frac{1}{{{\Lambda ^2}}}\left\{ {{{\left[ {{\theta ^2}{s^2}\left( {{\theta ^2} + 2{\eta ^2}} \right) + \left( {{\eta ^2}{s^2} + 1} \right)\left( {\eta  + s} \right)\left( {\eta  - s} \right)} \right]}^2} + 4{\theta ^2}{s^4}{{\left[ {{\theta ^2} + {\eta ^2} + 1} \right]}^2}} \right\} .  \label{R_(D)}
\end{equation}

\begin{figure}[h]
        \centering
        \begin{subfigure}[b]{0.5\textwidth}
                \includegraphics[width=\textwidth]{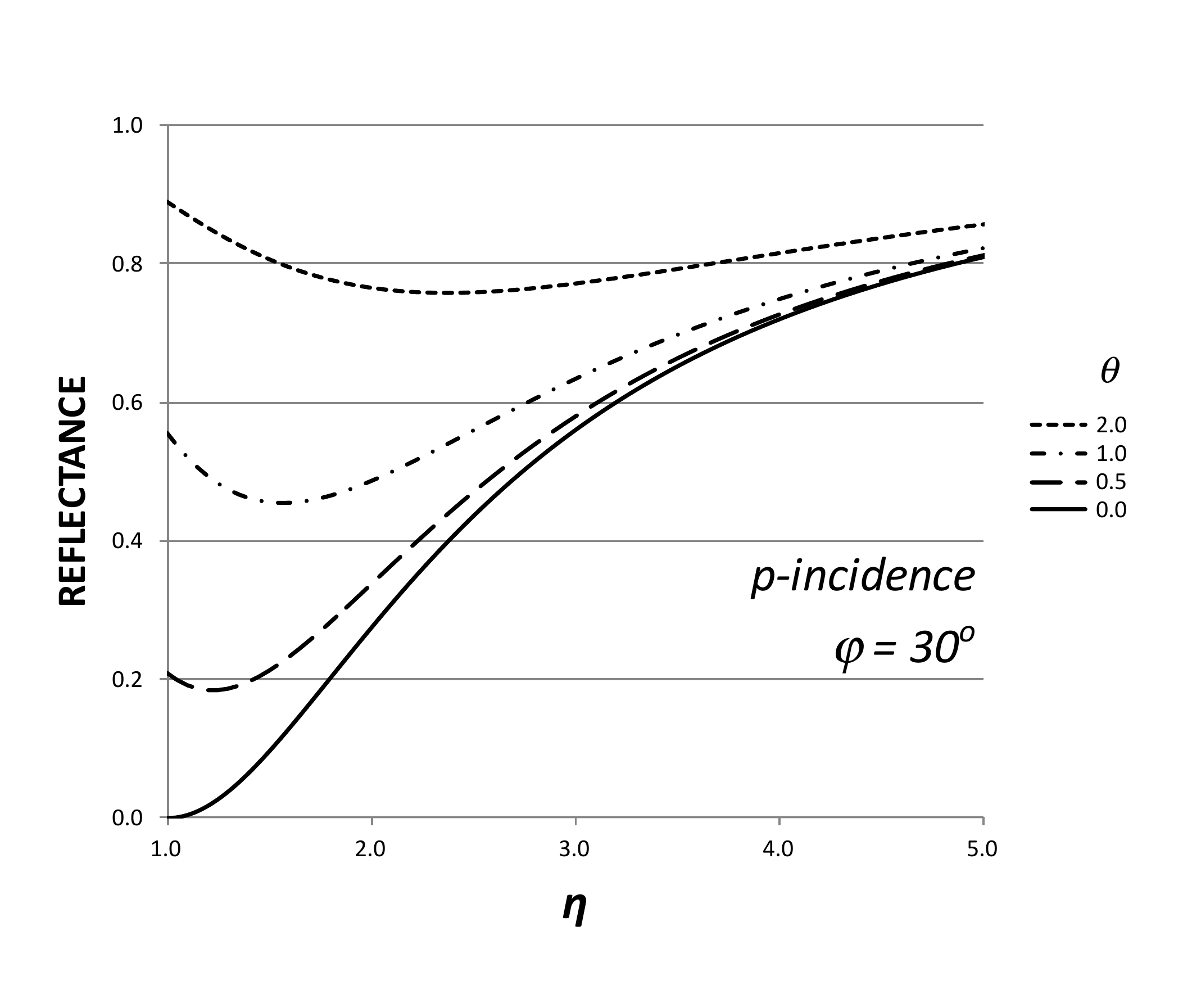}
                \caption{}
                \label{fig_FilmReflectance_eta-a}
        \end{subfigure}%
        ~ 
        \begin{subfigure}[b]{0.5\textwidth}
                \includegraphics[width=\textwidth]{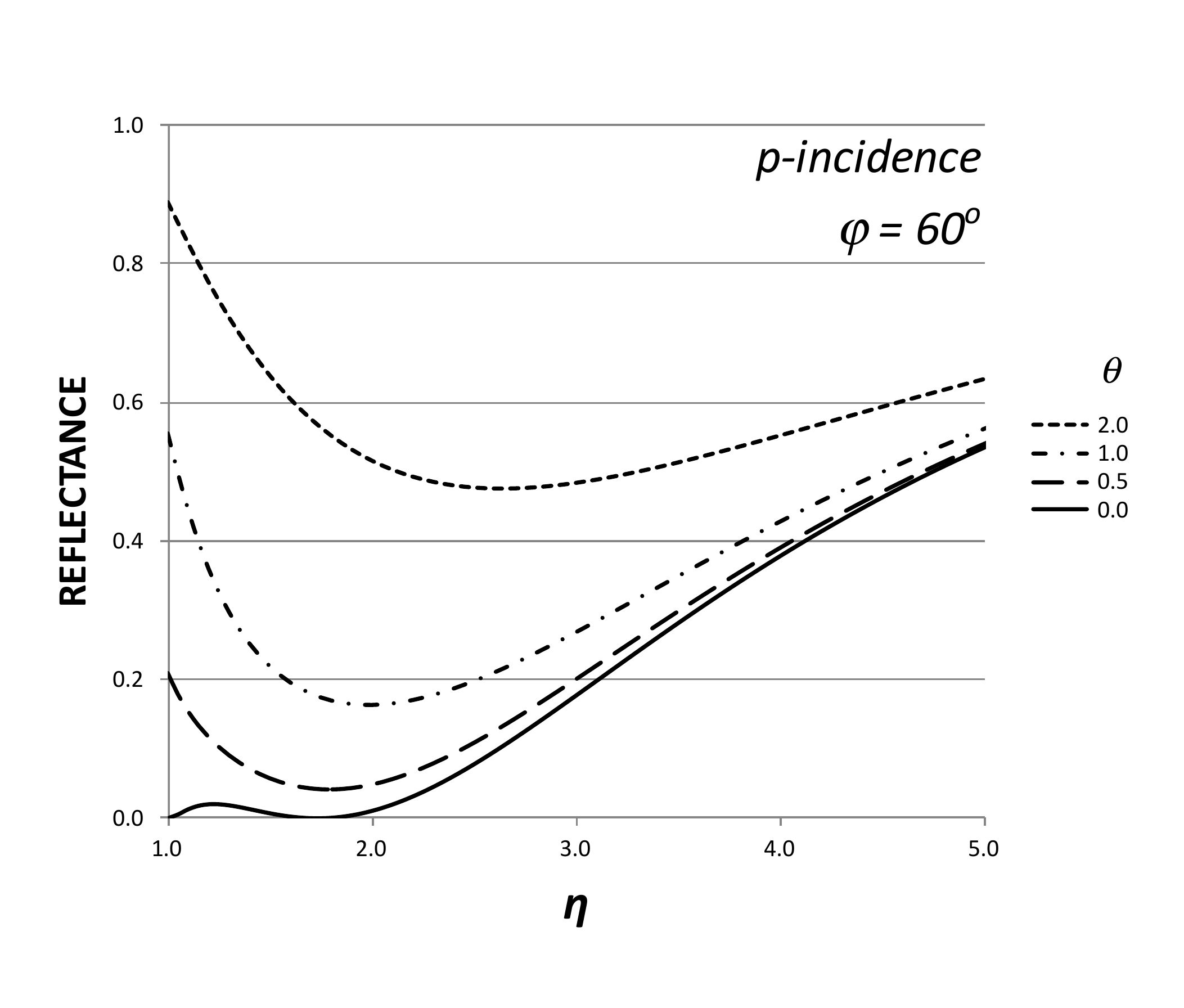}
                \caption{}
                \label{fig_FilmReflectance_eta-b}
        \end{subfigure}
        ~ 
               \caption{$\theta$-film reflectance, in the case of destructive interference and for two angles of incidence. The solid curve is the results for ordinary non-$\theta$ systems. In both cases, the $\theta$-film reflectance is much higher than for ordinary materials.}
               \label{fig_FilmReflectance_eta}
\end{figure}

Figure \ref{fig_FilmReflectance_eta} shows the $\theta$-film reflectance for two different angles of incidence on the film. The solid curve represents, in both graphs, the non-$\theta$, ordinary material results. Compared to ordinary materials, $\theta$-films are more opaque to radiation, particularly for a small $\eta$ where normal materials become more transparent. The larger $\eta$ is, the more reflective the film. But what is interesting here is that for topological materials the effect is enhanced, with an interplay that leads to a local minimum value in terms of $\eta$.

In Fig.~\ref{fig_FilmReflectance-theta-a} below, the curves for film reflectance versus $\theta$, and for different values of $\eta$ parameter, show the strong influence of the parameter $\theta$ in energy distribution (in the case of destructive interference). It is illustrative to note the curve for $\eta =1$ (solid curve), which represents the case where the medium inside the film and in the surroundings is the same, except for a nonvanishing value of $\theta$ in the film. This case can be also considered as representative of the pure vacuum case and it demonstrates the significatively different behavior of $\theta$-systems.

\begin{figure}[h]
        \centering
        \begin{subfigure}[b]{0.5\textwidth}
                \includegraphics[width=\textwidth]{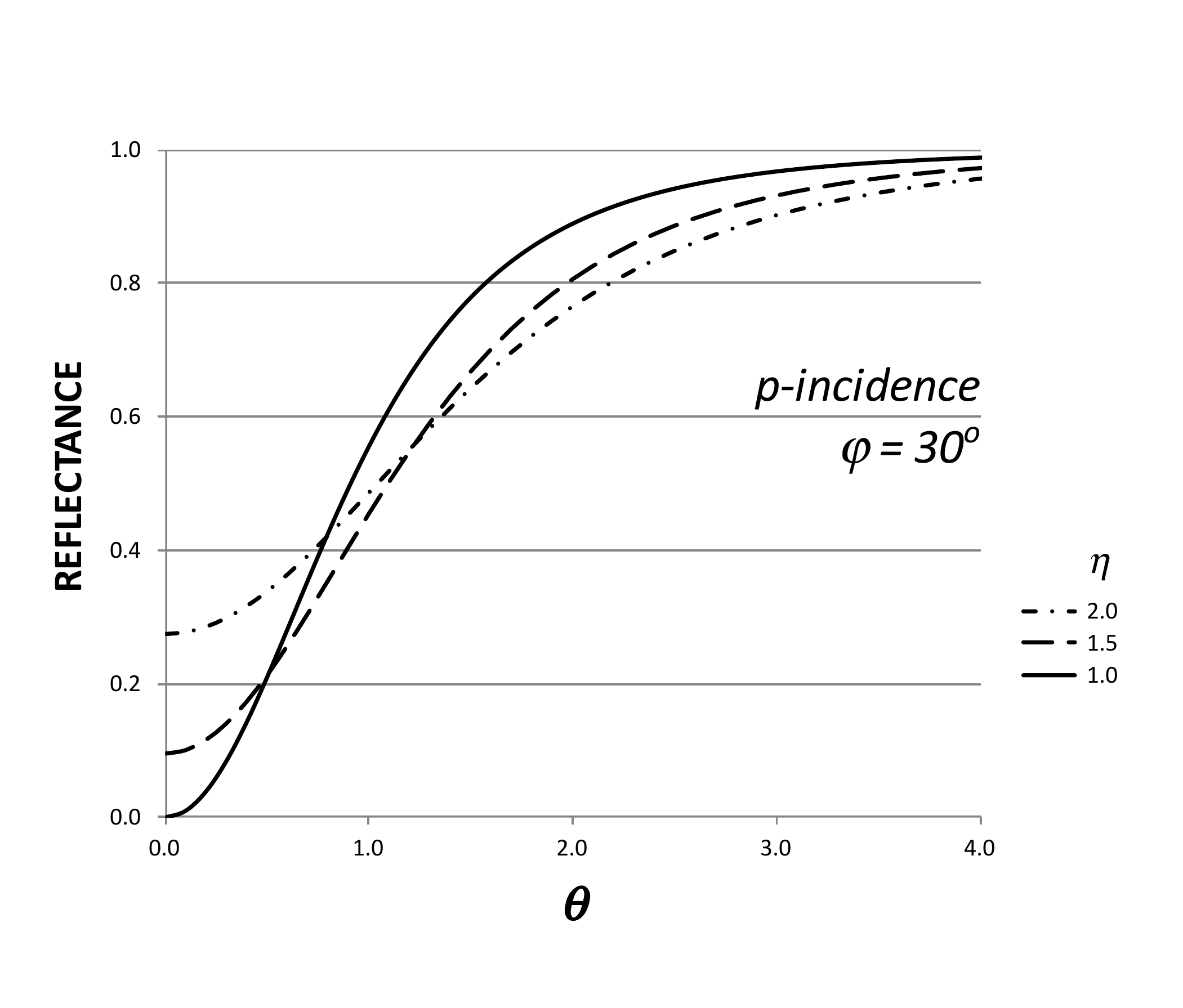}
                \caption{}
                \label{fig_FilmReflectance-theta-a}
        \end{subfigure}%
        ~ 
        \begin{subfigure}[b]{0.5\textwidth}
                \includegraphics[width=\textwidth]{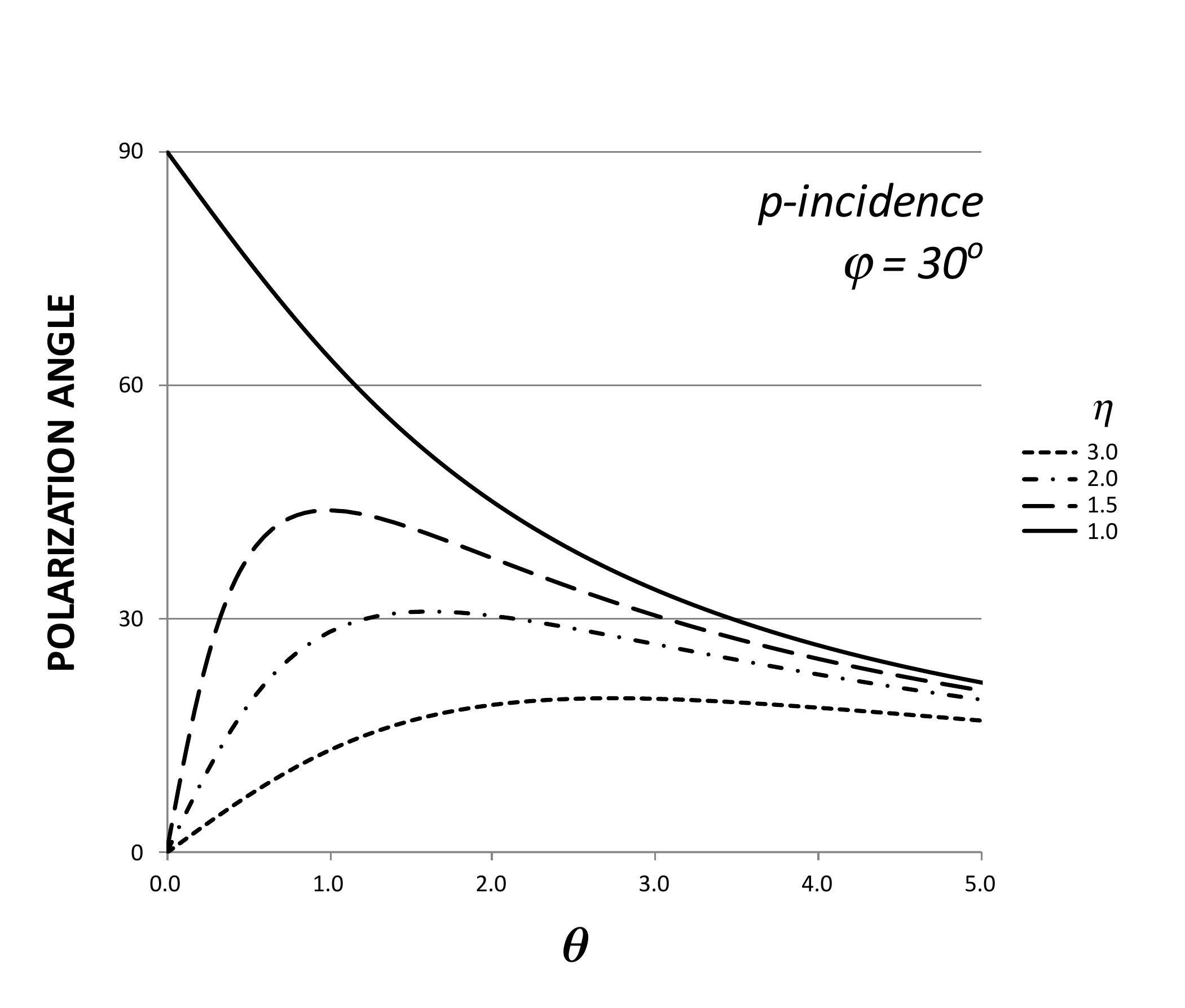}
                \caption{}
                \label{fig_FilmRef_Pol-theta-b}
        \end{subfigure}
        ~ 
               \caption{$\theta$-film (a) reflectance and (b) polarization of reflected radiation versus $\theta$, for destructive interference. The solid curve ($\eta = 1$) represents the case where the magnetoelectric properties are the same for the $\theta$-system and its surroundings (the full vacuum case is also included in such a situation).}
               \label{fig_FilmRef&Pol-theta}
\end{figure}

\section{POLARIZATION}

Only for destructive interference, is there a change in the radiation polarization. For a linear polarization of the incident wave, a polarization rotation appears for the reflected back and transmitted radiation by the film. For transmitted radiation beyond the film, we obtain, for the polarization angle $\alpha_T \equiv  \tan^{-1} \left( {\cal E}_{T \bot }/{\cal E}_{T\parallel } \right)$ ($\alpha_i$ being the incident wave polarization angle),
\begin{equation}
\alpha _T = \tan^{-1} \left[ \frac{{\theta s\left( {{s^2} - 1} \right) + \left[ {\left( {D - 2\eta {s^2}} \right) + \left( {\eta  + s} \right)\left( {{s^2} - 1} \right)} \right]\tan {\alpha _i}}}{{{s^2}\left( {D - 2\eta } \right) - s\left( {\eta s + 1} \right)\left( {{s^2} - 1} \right) + \theta s\left( {{s^2} - 1} \right)\tan {\alpha _i}}} \right] . \label{T-pol}
\end{equation}
For the reflected radiation, we find for the polarization angle $\alpha_R$,
\begin{equation}
\alpha _R = \tan^{-1} \left[ \frac{{2\theta s\left[ {D - \eta \left( {{s^2} + 1} \right)} \right] + \left[ { - {D^2} + 2\left[ {s\left( {\eta s + 1} \right) + \eta } \right]D - 2\eta s\left( {\eta s + 1} \right)\left( {{s^2} + 1} \right)} \right]\tan {\alpha _i}}}{{{{\left[ {D - \left( {s\left( {\eta s + 1} \right) + \eta } \right)} \right]}^2} - {\eta ^2}\left( {{s^4} - 1} \right) - {s^2} + 2\theta s\left[ {D - \eta \left( {{s^2} + 1} \right)} \right]\tan {\alpha _i}}} \right] . \label{R-pol}
\end{equation}

For p incidence, (\ref{R-pol}) and (\ref{T-pol}) become, respectively,
\begin{equation}
\alpha _T^{\left( p \right)} = \tan^{-1} \left[ \frac{{\theta \left( {{s^2} - 1} \right)}}{{s\left( {D - 2\eta } \right) - \left( {\eta s + 1} \right)\left( {{s^2} - 1} \right)}} \right]
\end{equation}
and
\begin{equation}
\alpha _R^{\left( p \right)} =  \tan^{-1} \left[ \frac{{2\theta s\left[ {{\theta ^2}s + s\left( {{\eta ^2} + 1} \right)} \right]}}{{{s^2}{{\left( {{\theta ^2} + {\eta ^2}} \right)}^2} - {\eta ^2}\left( {{s^4} - 1} \right) - {s^2}}} \right] .
\end{equation}

\begin{figure}[H]
        \centering
        \begin{subfigure}[b]{0.5\textwidth}
                \includegraphics[width=\textwidth]{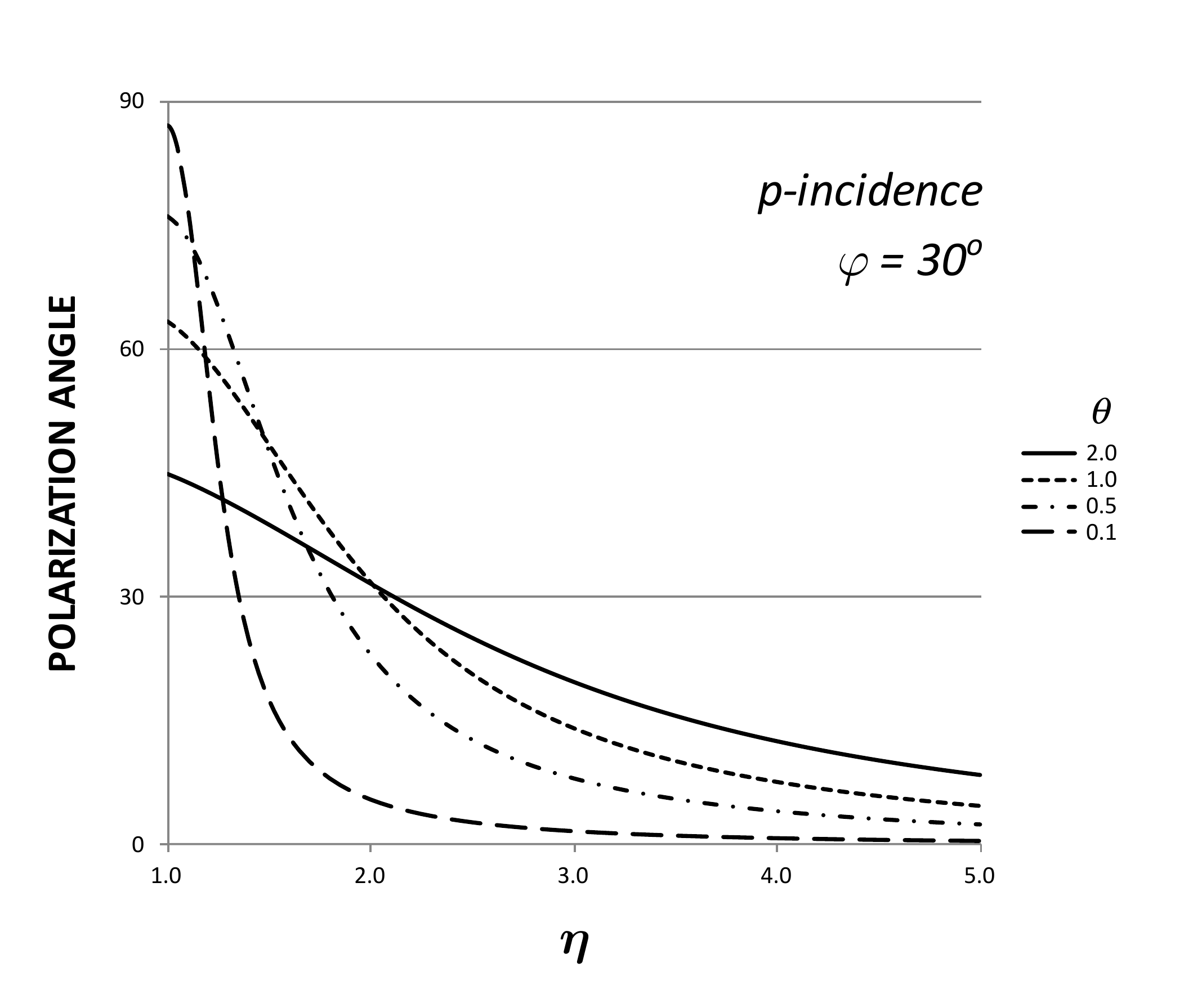}
                \caption{}
                \label{fig_FilmPolarization_eta-a}
        \end{subfigure}%
        ~ 
        \begin{subfigure}[b]{0.5\textwidth}
                \includegraphics[width=\textwidth]{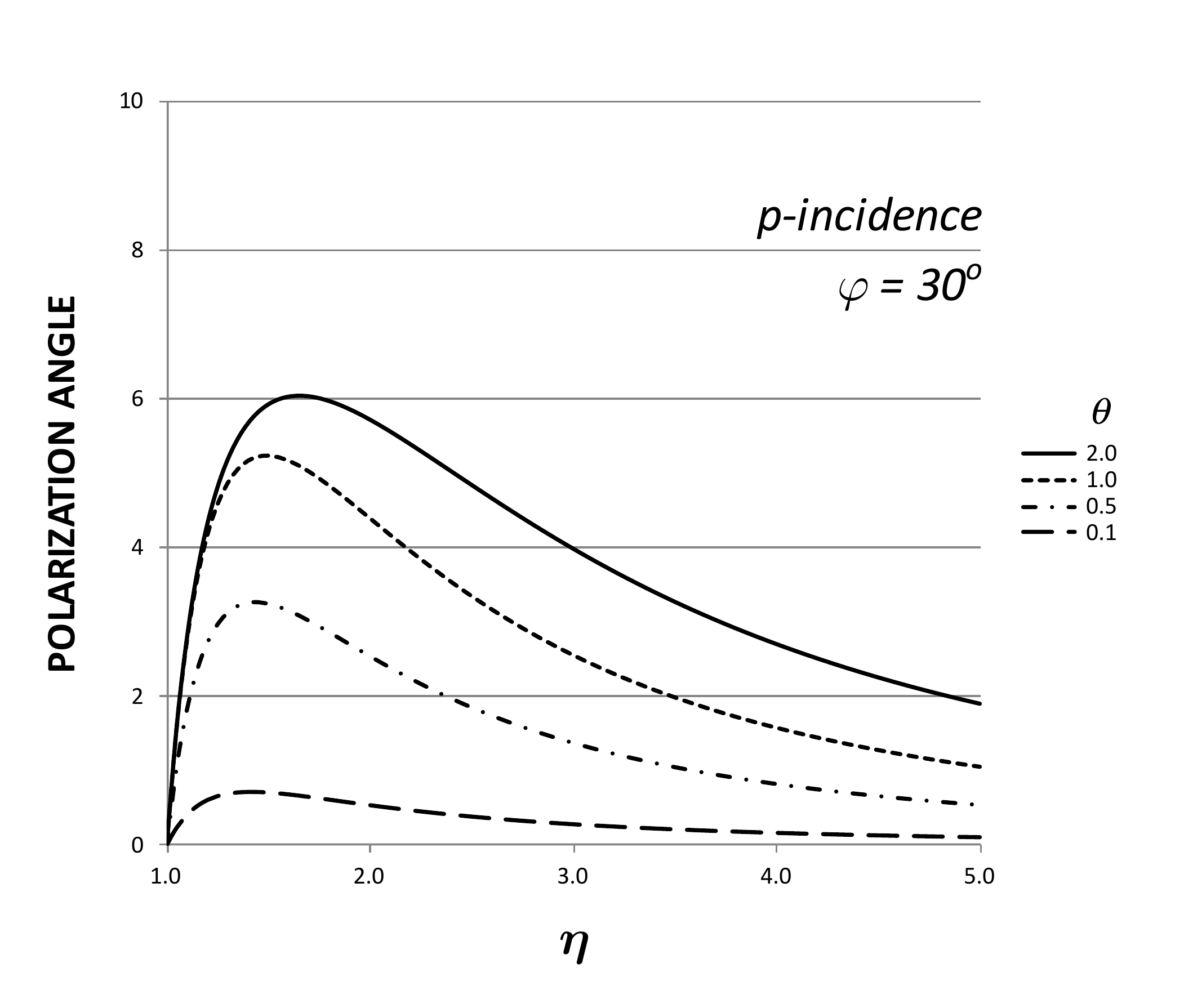}
                \caption{}
                \label{fig_FilmPolarization_eta-b}
        \end{subfigure}
        ~ 
               \caption{Polarization, in the case of destructive interference and for p-polarized incident radiation, for (a) the radiation reflected back and (b) the radiation transmitted through the film. The reflected radiation experiences a larger polarization rotation.}
               \label{fig_FilmPolarization_eta}
\end{figure}

In Fig.~\ref{fig_FilmRef_Pol-theta-b} above, we show the curves for film radiation polarization in terms of $\theta$ and for different values of $\eta$. In Fig.~\ref{fig_FilmPolarization_eta-a}, we see  the reflected radiation polarization, which undergoes a significant rotation. On the other hand, in Fig.~\ref{fig_FilmPolarization_eta-b}, we observe that the polarization of the radiation transmitted through the film is only slightly rotated (note the scale in the vertical axis). Moreover, a maximum for polarization rotation exists for both transmitted and reflected radiation, with a value of $\theta$ for that maximum that depends on the magnetoelectric properties and different for each wave. The closer the material to vacuum space, the more notorious the effect. Notice that, in this latter case, though polarization goes to $90^o$ as $\theta\to 0$, the reflected radiation polarization for $\theta =0$ actually vanishes.

\section{DISCUSSION}

A system with a Chern-Simons topology, being matter or vacuum, exhibits optical properties that differ from ordinary systems.  When the system is manufactured in the form of a thin film, interference effects are relevant, the film becomes iridescent and reflectance is enforced when the optical path across the film is consistent with destructive interference.  Actually, the larger is $\theta$ and the smaller is $\eta$, the greater the fraction of energy reflected back. The minimum transparency is achieved when the magnetoelectric properties are similar for the film and the surrounding medium ($\eta \sim 1$).

For the $\theta$-film in pure vacuum, the effects are quite strong. Film reflectance for destructive interference becomes
\begin{equation}
R^{({\rm D})}_{\scriptscriptstyle{\rm vacuum}}=\frac{\theta^2\left(\theta^2+4\right)}{\left(\theta^2+2\right)^2} ,
\end{equation}
significatively greater than the reflectance of a single interface [which is $\theta^2/(\theta^2+4)$]. A more extensive analysis of this, particularly for a multilayered configuration with different values of $\theta$ for each layer will be published elsewhere. A possible application of these results to astronomical issues is under investigation.

Although this paper is oriented to understand the optical behaviour of a generic $\theta$-film, where the CS term influences the magnetoelectric response to radiation, we consider the possibility that these results make sense for topological insulators. For this, the substitution $\theta/2 \to \theta e^2/4\pi^2$ must be assumed in the CS term of the action ($e$ is the dimensionless unit charge). Thus, the relevance of the present study for these systems is limited to small values of the parameter  $\theta$ used here. However, even for small $\theta$ the effects are significant. Of course, since we have considered a continuous domain of values for $\theta$, we are mainly aiming at a phase with broken time-reversal symmetry. In addition, the actual value of $\theta$ in topological insulators must also be investigated in terms of the band structure \cite{TI2008,Sekine2014}. In fact, $\theta$ behaves as a magnetoelectric polarizability \cite{EMV2009}, a feature also seen in our results.

\noindent
============================
\section*{\Large Acknowledgments}

I am very grateful to J. Zanelli for discussions and suggestions. This work was partially supported by CONICYT-PIA Grant No. ACT-1115.  The P4-Center for Research and Applications in Plasma Physics and Pulsed Power Technology is partially supported by the Comisi\'{o}n Chilena de Energ\'{\i}a Nuclear. I also greatly appreciate the close and constant support of Mariana Huerta.

\end{document}